\input mtexsis
\input epsf
\paper
\singlespaced
\widenspacing
\twelvepoint
\Eurostyletrue
\thicksize=0pt
\sectionminspace=0.1\vsize

\def\ga{g_{\scriptscriptstyle A}}
\def\nc{{N_c}}
\def\yop{{f_\pi}}
\def\yo2{{f_\pi^2}}

\def\oneht{\textstyle{1\over 2} }
\def\onehtss{\scriptscriptstyle{1\over 2} }
\def\threehtss{\scriptscriptstyle{3\over 2} }

\def\threeht{\textstyle{3\over 2} }

\def\curlyR{{\cal R}}
\def\sss{\scriptscriptstyle}


\referencelist

\reference{burkert}
For a recent review, see V.D.~Burkert, {{\tt hep-ph/0210321}}
\endreference


\reference{latt}
S.~Sasaki, T.~Blum and S.~Ohta, 
\journal Phys. Rev. D.;65,074503 (2002)
\endreference

\reference{*latta} M.~Gockeler {\it et al} [QCDSF Collaboration], {{\tt nucl-th/0206049}}
\endreference

\reference{*lattb}
D.~Richards, 
\journal Nucl. Phys. B (Proc. Suppl.);94, 269 (2001)
\endreference

\reference{*lattc}
W.~Melnitchouk {\it et al},
\journal Nucl. Phys. Proc. Suppl.;109, 96 (2002)
\endreference

\reference{*lattd}
W.~Melnitchouk {\it et al}, {{\tt hep-lat/0210042}}
\endreference

\reference{*latte}
F.X.~Lee, D.B.~Leinweber, \journal Nucl. Phys. B (Proc. Suppl.);73, 258 (1999)
\endreference

\reference{*lattf}
F.X.~Lee {\it et al}, \journal Nucl. Phys. B (Proc. Suppl.);106, 248 (2002)
\endreference


\reference{largen} C.D.~Carone {\it et al},
\journal Phys. Rev. D;50, 5793  (1994) 
\endreference

\reference{*largena}
J.L.~Goity, \journal Phys. Lett. B;414,140 (1997)
\endreference

\reference{*largenb}
C.E.~Carlson and C.D.~Carone, \journal Phys. Rev. D;58, 053005 (1998)
\endreference

\reference{*largenc}
C.E.~Carlson  {\it et al},
\journal Phys. Lett. B;438,327 (1998); \journal Phys. Rev. D;59,114008 (1999)
\endreference

\reference{*largend}  D.~Pirjol and T-M.~Yan,
\journal Phys. Rev. D;57,1449 (1998); {\it ibid}, 5434
\endreference

\reference{*largene}
C.L.~Schat, J.L.~Goity and N.N.~Scoccola, {{\tt hep-ph/0111082}}; {{\tt hep-ph/0209174}}
\endreference

\reference{Carlson:2000zr}
C.E.~Carlson and C.D.~Carone, \journal Phys. Lett. B;484,260 (2000)
\endreference

\reference{Manohar:1998xv}
For a review, see A.V.~Manohar, {{\tt hep-ph/9802419}}
\endreference

\reference{Isgur}
N.~Isgur and G.~Karl, \journal Phys. Rev. D;19,2653 (1979)
\endreference

\reference{riska}
L.Y.~Glozman and D.O.~Riska,
\journal Phys. Rept.;268, 263 (1996)
\endreference

\reference{brown}
I.~Zahed and G.E.~Brown,
\journal Phys. Rept.;142, 1 (1986)
\endreference

\reference{Lee:2002gn}
F.X.~Lee {\it et al}, {{\tt hep-lat/0208070}}
\endreference

\reference{Li:1991yb}
Z.p.~Li, V.~Burkert and Z.j.~Li, 
\journal Phys. Rev. D;46,70 (1992)
\endreference

\reference{Carlson:1991tg}
C.E.~Carlson and N.C.~Mukhopadhyay,
\journal Phys. Rev. Lett.;67, 3745 (1991)
\endreference

\reference{Tom}
T.D.~Cohen and L.Y.~Glozman,
\journal Phys. Rev. D;65, 016006 (2002)
\endreference

\reference{oka}
D.~Jido, M.~Oka and A.~Hosaka,
\journal Prog. Theor. Phys.;106, 873 (2001)
\endreference


\reference{Harari}
H.~Harari, \journal Phys. Rev. Lett.;16, 964 (1966)
\endreference

\reference{*Hararia}
H.J.~Lipkin, H.R.~Rubinstein and S.~Meshkov,
\journal Phys. Rev.;148, 1405 (1966)
\endreference

\reference{*Hararib}
I.S.~Gerstein and B.W.~Lee,
\journal Phys. Rev. Lett.;16, 1060 (1966)
\endreference

\reference{*Hararic}
H.~Harari,
\journal Phys. Rev. Lett.;17, 56 (1966)
\endreference

\reference{Fred}
F.~Gilman and H.~Harari,
\journal Phys. Rev.;165, 1803 (1968)
\endreference

\reference{wein1}  S.~Weinberg, \journal Phys. Rev.;177,2604 (1969)
\endreference


\reference{strong}  
S.~Weinberg, in {\it Chiral dynamics: theory and experiment},
edited by A.M.~Bernstein and B.R.~Holstein, (Springer-Verlag, 1995), {{\tt hep-ph/9412326}}.
\endreference

\reference{alg} B.~Zumino, in {\it Theory and Phenomenology in Particle
Physics}, 1968 International School of Physics `Ettore Majorana'
(Academic Press, New York, 1969) p.42
\endreference

\reference{*alga} S.~Weinberg, in {\it Lectures on Elementary Particles and
Quantum Field Theory}, edited by Stanley Deser {\it et al}, 
(MIT Press, Cambridge, MA, 1970) p.285
\endreference 

\reference{*algb}  For a review, see R.~de Alfaro, S.~Fubini, G.~Furlan, and G.~Rossetti, 
            {\it Currents in Hadron Physics}, (North-Holland, Amsterdam, 1973)
\endreference

\reference{adler} S.L.~Adler, \journal Phys. Rev. Lett.;14,1051 (1965) 
\endreference

\reference{*adlera} W.I.~Weisberger, \journal Phys. Rev. Lett.;14,1047 (1965) 
\endreference

\reference{pdg} K.~Hagiwara {\it et al}  [Particle Data Group],
\journal Phys. Rev. D;66,010001 (2002)
\endreference

\reference{beane2}  S.R.~Beane, \journal Phys. Rev. D;59,031901(1999)
\endreference

\reference{Fettes}  
N.~Fettes,
JUL-3814 (PhD Thesis)
\endreference

\reference{beanesav}  S.R.~Beane and M.J.~Savage, {\it in preparation}
\endreference

\reference{Casher}
A.~Casher and L.~Susskind,
\journal Phys. Rev. D;9, 436 (1974)
\endreference

\endreferencelist

\titlepage
\obeylines
\hskip4.8in{INT-PUB-02-52}\unobeylines
\vskip0.7in
\title

The Role of the Roper in QCD

\endtitle
\vskip0.2in

\author
S.R.~Beane$^{\, a}$ and U.~van Kolck$^{\, b,c}$

\vskip0.01in

{\it $^{\rm a}$Institute for Nuclear Theory,  
University of Washington, Seattle, WA 98195-1560}

\vskip0.1in

$^b${\it Department of Physics, 
University of Arizona, Tucson, AZ 85721}

\vskip0.1in

$^c${\it RIKEN-BNL Research Center, 
Brookhaven National Laboratory, Upton, NY 11973}

\endauthor

\vskip0.1in

\abstract
\singlespaced
\widenspacing

We show that existing data suggest a simple scenario in which the nucleon, and
the $\Delta$ and Roper resonances act as chiral partners in a reducible
representation of the full QCD chiral symmetry group. We discuss the peculiar
interpretation of this scenario using spin-flavor symmetries of the naive
constituent quark model, as well as the consistency of the scenario with
large-$\nc$ expectations.

\endabstract 
\vskip0.5in
\center{PACS: 11.30.Rd; 12.38.Aw; 11.55.Jy; 11.30.Er} 
\endcenter
\endtitlepage
\vfill\eject                                     
\superrefsfalse
\singlespaced
\widenspacing

\vskip0.1in
\noindent {\twelvepoint{\bf 1.\quad Introduction}}
\vskip0.1in

\noindent Understanding of the pattern of baryon masses and couplings from QCD
remains an open challenge for theorists. The need for theoretical progress is
at least partly driven by the prospect of new experimental results: the CLAS
collaboration at JLab expects to significantly improve knowledge of excited
baryon masses and decays\ref{burkert}. Complementary efforts are well underway to
compute properties of excited baryons using lattice QCD\ref{latt}. Progress
has also been made in analyzing the excited baryons in the $1/\nc$
expansion\ref{largen}\ref{Carlson:2000zr}.  In the large-$\nc$ limit, the light
baryons fall into representations of a contracted $SU(2 N_f)$ spin-flavor
symmetry where $N_f$ is the number of active flavors\ref{Manohar:1998xv}. This
symmetry is highly predictive; it determines {\it inter alia} ratios of
axial-vector couplings and magnetic moments, and gives rigorous justification
to several aspects of the naive constituent quark model (NCQM).

One particularly interesting unresolved issue in baryon spectroscopy is that of
the role of the Roper resonance, $N(1440)$ or $N'$. In the most naive
interpretation, $N'$ is a three-quark radial excitation of the nucleon with the
same spin-parity quantum numbers. However, this interpretation has been
questioned for several reasons. First, the calculated mass of $N'$ would appear
to be too high in quark models which include one-gluon
exchange\ref{Isgur}. 
(Models with explicit pion degrees of freedom evidently do not suffer from this drawback. See, for instance,
\Ref{riska} and \Ref{brown}.)
Second, several recent quenched lattice QCD calculations
find a spectrum inverted$^1$\vfootnote1{\tenpoint See, however, a recent study
which uses Bayesian techniques\ref{Lee:2002gn}.} with respect to experiment,
with $N'$ heavier than the first excited state with opposite
parity\ref{latt}. An alternative interpretation is that the Roper is a
hybrid state; that is, it couples predominantly to QCD currents with some
gluonic contribution\ref{Li:1991yb}\ref{Carlson:1991tg}. Recently, the
consequences of spin-flavor $SU(6)$ for the Roper multiplet have been worked
out in the large-$\nc$ expansion, assuming that the Roper is in a 
${\bf 56}$-dimensional representation of spin-flavor $SU(6)$ 
(the ${\bf 20}$-dimensional representation of spin-flavor
$SU(4)$)\ref{Carlson:2000zr}. Several predictions have been made which will 
be tested experimentally at Jlab and other experimental facilities.

In this paper we consider consequences of chiral symmetry for the
low-lying light baryons. When chiral constraints on the baryons are discussed it is almost
always in the context of the limiting scenario of 
chiral-symmetry restoration$^2$\vfootnote2{\tenpoint For several recent
attempts, see \Ref{Tom} and \Ref{oka}.}.
By contrast, we will consider consequences of the full QCD chiral symmetry
group in the broken phase for the light baryons. An important message that this
paper hopes to convey is that there is no need to summon extreme conditions
in order to find consequences of chiral symmetry in the broken chiral symmetry
phase. The formalism necessary to extract the consequence of chiral
symmetry for baryons was developed by Weinberg and others many years
ago\ref{Harari}\ref{Fred}\ref{wein1}\ref{strong}\ref{alg}.
We find it surprising that this ancient wisdom is not widely known. 
Thus while much of what we present in this paper is not new, we feel
that the time is ripe for a reassessment of these powerful methods.

We first motivate our discussion by performing an updated analysis of the
well-known Adler-Weisberger sum rule for pion-nucleon ($\pi N$)
scattering\ref{adler}. We then show that in the resonance-saturation
approximation, the Adler-Weisberger sum rule can be derived directly from the
chiral $SU(2)\times SU(2)$ algebra through a simple application of the
Wigner-Eckart theorem. Not surprisingly, the most powerful way of deriving
consequences of the chiral algebra is by using the group representation theory:
that is, by requiring that the baryons transform as sums of allowed irreducible
representations. Now, naively one might expect that these representations are
infinite dimensional and that consequently chiral symmetry gives very little
predictive power in the low-energy theory. However, as we will show, data
suggest otherwise. In particular, we will make the case that the ground-state
chiral multiplet is composed of the nucleon, $N(940)$, the $\Delta$ resonance,
$\Delta (1232)$, and the Roper resonance, $N(1440)$, which fall into a
reducible $(0, \oneht )\oplus (\oneht ,1)$ representation of $SU(2)\times
SU(2)$ with approximately maximal mixing. This representation offers a
compelling interpretation of the nucleon axial couplings  and of the special role
of the Roper resonance in QCD.

We offer an interpretation of our results in the context of the
NCQM. We find that our proposed chiral
representation is equivalent to placing the nucleon and the $\Delta$
and Roper resonances in a reducible ${\bf 4}\oplus{\bf 20}$ representation of
spin-flavor $SU(4)$. While this scenario is consistent with
large-$\nc$ QCD, the naive large-$\nc$ counting is badly violated by
experiment. Moreover, we find that our results are not consistent with placing
the Roper in a ${\bf 20}$-dimensional representation of spin-flavor $SU(4)$, as
is usually assumed; rather, consistency of our results with large $\nc$ would require
that the Roper be in the fundamental representation of $SU(4)$ in the
large-$\nc$ limit.

\vskip0.1in
\noindent {\twelvepoint{\bf 2.\quad The Adler-Weisberger Sum Rule}}
\vskip0.1in

\vskip0.1in
\noindent {\twelvepoint{\it 2.1\quad  The Dispersion Relation}}

\noindent Consider the renowned Adler-Weisberger sum rule\ref{adler},

$$
\ga^2=1-{{2\yo2}\over\pi}\int_0^\infty {{d\nu}\over\nu}
\lbrack\sigma^{\pi^- p}(\nu )-\sigma^{\pi^+ p}(\nu )\rbrack .
\EQN awsumrule$$
Here $\ga$ is the nucleon axial-vector coupling, ${\yop}\simeq 93~{\rm MeV}$ is
the pion decay constant and $\sigma^{\pi^\pm p}$ is the total cross-section for
charged pion scattering on a proton.
Recall that this sum rule for the $\pi N$ scattering amplitude follows from two
inputs: (i) a chiral symmetry low-energy theorem and (ii) the assumption that
the forward $\pi N$ amplitude with isospin, $I=1$, in the $t$-channel satisfies 
an unsubtracted dispersion relation. Saturating the sum rule with $N$ ($I=1/2$) and $\Delta$
($I=3/2$) resonances gives

$$
\ga^2=1-\sum_{\sss N} {\cal I}_{\sss N}+\sum_{\sss \Delta} {\cal I}_{\sss \Delta}
+\, {\it continuum},
\EQN awsumrulepolesat$$
where the ${\cal I}_\curlyR$ are related to experimental widths by

$$
{\cal I}_\curlyR ={{{64\pi\yo2}{M_\curlyR^3}}\over{3(M_\curlyR^2-M_N^2)^3}}
\left(S_\curlyR+\frac{1}{2}\right)
\Gamma^{\scriptstyle {\rm TOT}}(\curlyR\rightarrow N\pi ),
\EQN sumruleinput$$
and $S_\curlyR$ is the spin of the resonance $\curlyR$.  

We can now go to the
Particle Data Group (PDG)\ref{pdg} and compute the contribution of each $N$ and
$\Delta$ state to the sum rule (see Table 1). 
We include only established resonances 
($\star\star\star$ and $\star\star\star\;\star$), using 
PDG central values and
estimates. We find $\sum {\cal I}_N=0.72$ and $\sum {\cal I}_\Delta =1.3$.  Neglecting the continuum
contribution (we will return to this point below), 
we then obtain $\ga =1.26$, to be compared to the
experimental value of $1.2670\pm 0.0035$\ref{pdg}. This is truly remarkable
agreement.  There are several important things to notice from Table 1.
First, there is a cancellation between the $N$- and $\Delta$-type
contributions, which enter with opposite sign.
Second,
$\Delta (1232)$ and $N(1440)$ dominate the sum rule.
Axial transitions of the excited baryons to the ground-state nucleon
are small compared to the dominant transitions. 
For instance, saturating the sum rule with these two states alone gives $\ga=1.34$.

\table{gl}
\tenpoint
\caption{\twelvepoint Resonances which contribute to the Adler-Weisberger sum
rule for $\pi N$ scattering. 
We have used PDG central 
values and
estimates. We emphasize that there is substantial uncertainty in these values.
Only established resonances 
($\star\star\star$ and $\star\star\star\;\star$) have been tabulated.}
\doublespaced
\ruledtable
{}    | $\curlyR$  | ${\cal I}_\curlyR$ \dbl ${}$  | $\curlyR$ | ${\cal I}_\curlyR$  \cr
$P_{11}\;({1\over 2}^+)$|$N(940)$ | $\>\;\; --$ \dbl $P_{11}\;({1\over 2}^+)$ | $N(1710)$ | $\>\;\;0.01$ \cr
$P_{33}\;({3\over 2}^+)$|$\Delta (1232)$ | $\>\;\;1.02$ \dbl $P_{13}\;({3\over 2}^+)$ | $N(1720)$ | $\>\;\;0.02$ \cr
$P_{11}\;({1\over 2}^+)$|$N(1440)$  | $\>\;\;0.23$ \dbl $F_{35}\;({5\over 2}^+)$ | $\Delta (1905)$ | $\>\;\;0.02$ \cr
$D_{13}\;({3\over 2}^-)$|$N(1520)$  | $\>\;\;0.09$ \dbl $P_{31}\;({1\over 2}^+)$ | $\Delta (1910)$ | $\>\;\;0.01$ \cr
$S_{11}\;({1\over 2}^-)$|$N(1535)$  | $\>\;\;0.04$ \dbl $P_{33}\;({3\over 2}^+)$ | $\Delta (1920)$ | $\>\;\;0.01$ \cr
$P_{33}\;({3\over 2}^+)$|$\Delta (1600)$  | $\>\;\;0.06$ \dbl $D_{35}\;({5\over 2}^-)$| $\Delta (1930)$ | $\>\;\;0.03$ \cr
$S_{31}\;({1\over 2}^-)$|$\Delta (1620)$  | $\>\;\;0.02$ \dbl $F_{37}\;({7\over 2}^+)$| $\Delta (1950)$ | $\>\;\;0.08$ \cr
$S_{11}\;({1\over 2}^-)$|$N(1650)$  | $\>\;\;0.04$ \dbl $G_{17}\;({7\over 2}^-)$ | $N(2190)$ | $\>\;\;0.03$ \cr
$D_{15}\;({5\over 2}^-)$|$N(1675)$  | $\>\;\;0.08$ \dbl $H_{19}\;({9\over 2}^+)$ | $N(2220)$ | $\>\;\;0.03$ \cr
$F_{15}\;({5\over 2}^+)$|$N(1680)$  | $\>\;\;0.10$ \dbl $G_{19}\;({9\over 2}^-)$ | $N(2250)$ | $\>\;\;0.02$ \cr
$D_{13}\;({3\over 2}^-)$|$N(1700)$|$\>\;\;0.01$ \dbl $H_{3,11}\;({{11}\over 2}^+)$|$\Delta (2420)$ | $\>\;\;0.02$ \cr
$D_{33}\;({3\over 2}^-)$|$\Delta (1700)$|$\>\;\;0.03$ \dbl $I_{1,11}\;({{11}\over 2}^-)$ | $N(2600)$ | $\>\;\;0.02$  
\endruledtable
\endtable

Given the uncertainties in the resonance masses and axial couplings, and the
neglect of the continuum contribution, such remarkable agreement must to some
degree be fortuitous. Given the success of the sum rule one might ask: what
precisely is the sum rule testing about QCD? What is the significance of the
assumption about the asymptotic behavior of the forward $\pi N$ scattering
amplitude? Why do $\Delta (1232)$ and $N(1440)$ seem to have special status in
saturating the sum rule?  In order to answer these questions we will rephrase
the discussion of the sum rule entirely in the language of chiral symmetry.

\vskip0.1in
\noindent {\twelvepoint{\it 2.2\quad The Symmetry Interpretation}}

\noindent In the limit of vanishing up and down 
quark masses, QCD has an $SU(2)_L\times SU(2)_R$ invariance. We can write the chiral algebra as

\offparens
$$
[{{\cal Q}^{\sss A}_{a}},{{\cal Q}^{\sss A}_{b}}]=i{\epsilon_{abc}}{T_c}; 
\qquad
[{T^{\;\;}_a},{{\cal Q}^{\sss A}_{b}}]=
i{\epsilon_{abc}}{{\cal Q}^{\sss A}_{c}};
\qquad
[{T^{\;\;}_a},{T^{\;\;}_b}]=i{\epsilon_{abc}}{T^{\;\;}_c},
\EQN chiralalg
$$\autoparens 
where ${T^a}$ are $SU(2)_V$ generators and ${{\cal Q}^{\sss A}_{a}}$ are the
remaining axial generators.
We define the axial-vector coupling matrix,

\offparens
$$
\bra{\beta , \lambda}{{\cal Q}^{\sss A}_a}\ket{\alpha , \lambda'}=
[{X_a^\lambda}]_{\sss\beta\alpha}\,\delta_{\sss{\lambda\lambda'}},
\EQN xdefined$$\autoparens
where $\ket{\alpha , \lambda}$ is a baryon state of definite helicity
$\lambda$. Notice the Kronecker delta on the right side of this
equation.  This implies that we are defining ${X_a^\lambda}$ in a
helicity-conserving Lorentz frame\ref{wein1}.  A frame in which all momenta are collinear
is such a frame, as is the infinite-momentum frame. We are of course free
to choose any frame. However, helicity-conserving frames are special because in
these frames chirality conservation becomes the same as helicity conservation. 

The physical consequences of the full chiral symmetry group can now be
found by taking matrix elements of the $SU(2)\times SU(2)$ algebra and
using the Wigner-Eckart theorem to express the algebra as a set of
equations for reduced matrix elements\ref{wein1}. 
Taking matrix elements of the $SU(2)\times SU(2)$
algebra of \Eq{chiralalg} and inserting a complete set of states gives

\offparens
$$
[{X_a^\lambda},{X_b^\lambda}]_{\sss\beta\alpha}=
i{\epsilon_{abc}}[{T_c}]_{\sss\beta\alpha}.
\EQN aw$$\autoparens
This is a (generalized) Adler-Weisberger sum rule. An important comment is in
order here. One might suspect that the vacuum should contribute in the sum over states and
that the axial generator acting on the vacuum will generate quark-antiquark
pairs, thus destroying the group algebraic structure. The advantage of working in
a helicity-conserving frame is that the vacuum does not contribute in the sum
over states; i.e. ${{\cal Q}^{\sss A}_{a}}\ket{0}=0$$^3$\vfootnote3{
\tenpoint Note that the derivation of \Eq{aw} given in \Ref{alg} avoids any discussion of the
QCD vacuum.}.
The chiral symmetry is, however, broken 
spontaneously: although ${X_a^\lambda}$ satisfies the chiral
algebra, it does not commute with the baryon mass-squared matrix and is therefore not a
symmetry generator. Hence in helicity-conserving frames, all evidence of
symmetry breaking is in the Hamiltonian and not in the states.
Using the Wigner-Eckart theorem, we can write

$$
\bra{{I_\beta},{m_\beta}}{X_{(m)}^\lambda}\ket{{I_\alpha},{m_\alpha}}=
{{C}_{{I_\alpha}1}}({I_\beta},{m_\beta};{m_\alpha},{m})
X^\lambda ({\beta},{\alpha}),
\EQN we$$
where the $C$'s are Clebsch-Gordan coefficients and the $X^\lambda
({\beta},{\alpha})$ are reduced matrix elements\ref{wein1}.  Taking matrix
elements of \Eq{aw} between states of definite isospin and inserting a
complete set of states then gives a set of coupled equations for the
reduced matrix elements\ref{alg}. It is easy to show that for
$I_\alpha =I_\beta=\oneht$, \Eq{aw} yields

$$
\sum_{\sss \gamma =N} X^\lambda ({\beta},{\gamma})  X^{\lambda\dagger} ({\gamma},{\alpha})
-\sum_{\sss \gamma =\Delta} X^\lambda ({\beta},{\gamma})  X^{\lambda\dagger} ({\gamma},{\alpha})=
{\textstyle{3\over 4}}\,\delta_{\alpha \beta}.
\EQN aw2$$

The coupling of most interest here is 

$$
\ga=
\sqrt{2}\bra{\, p\,}\,{{X_{\sss{(+)}}^{\sss{\pm{1\over 2}}}}}\,\ket{\, n\,}=
\sqrt{\textstyle{4\over 3}}{X^{\sss{\pm{1\over 2}}}(\, p\,,\, n\,)},
\EQN onega $$
where $X_{\sss{(\pm)}}\equiv\mp\textstyle{1\over\sqrt{2}}(X_{\sss 1}\pm i X_{\sss 2})$
and we are ignoring overall phases. {}From \Eq{aw2} and \Eq{onega}, 
with $\alpha=\beta=N$ and $\gamma=\curlyR$, it immediately follows that

$$
\ga^2=1-{\textstyle{4\over 3}}
\sum_{\sss \curlyR =N} X^{\sss{1\over 2}}({N},{\curlyR})
X^{{\sss{1\over 2}}\dagger} ({\curlyR},{N})
+{\textstyle{4\over 3}}
\sum_{\sss \curlyR =\Delta} X^{\sss{1\over 2}}({N},{\curlyR}) 
X^{{\sss{1\over 2}}\dagger} ({\curlyR},{N}),
\EQN aw3$$
which is none other than \Eq{awsumrulepolesat} when we identify

$$
{\cal I}_\curlyR={\textstyle{4\over 3}}\, X^{\sss{1\over 2}}({N},{\curlyR})  
X^{{\sss{1\over 2}}\dagger} ({\curlyR},{N}).
\EQN identify$$

We have now derived the Adler-Weisberger sum rule in two ways. This is not surprising:
in \Ref{alg}, Weinberg proved that the assumption that the forward
$\pi N$ amplitude with $I=1$ in the $t$-channel satisfies an
unsubtracted dispersion relation is equivalent to the algebraic
statement of \Eq{aw} in helicity-conserving frames.  
In his original derivation, Weinberg used Regge pole theory to derive the
asymptotic constraint, which then led to
\Eq{aw}.

Note that, as a bonus, we can derive from \Eq{aw2}
relations for other couplings of interest.
These couplings involve, beside the ground-state nucleon $N$,
also the lowest excitations
such as the $\Delta (1232)$ and the $N(1440)$.
For example, we can define

$$\EQNalign{
&\ga'=
\sqrt{2}\bra{\, p\,}\,{{X_{\sss{(+)}}^{\sss{\pm{1\over 2}}}}}\,\ket{\, n'\,}=
\sqrt{\textstyle{4\over 3}}{X^{\sss{\pm{1\over 2}}}(\, p\,,\, n'\,)},
\EQN onega2;a \cr
&\ga''=
\sqrt{2}\bra{\, p'\,}\,{{X_{\sss{(+)}}^{\sss{\pm{1\over 2}}}}}\,\ket{\, n'\,}=
\sqrt{\textstyle{4\over 3}}{X^{\sss{\pm{1\over 2}}}(\, p'\,,\, n'\,)},
\EQN onega2;b \cr
&{\cal C}_{\sss\Delta N}=
\sqrt{3}\bra{\, p\,}\,{{X_{\sss{(+)}}^{\sss{\pm{1\over 2}}}}}
\,\ket{\,\Delta^{++}\,}=
\sqrt{\textstyle{3}}{X^{\sss{\pm{1\over 2}}}(\, p\,,\,\Delta^{++}\,)},
\EQN onega2;c \cr
&{\cal C}_{\sss\Delta N'}=
\sqrt{3}\bra{\, p'\,}\,{{X_{\sss{(+)}}^{\sss{\pm{1\over 2}}}}}
\,\ket{\,\Delta^{++}\,}=
\sqrt{\textstyle{3}}{X^{\sss{\pm{1\over 2}}}(\, p'\,,\,\Delta^{++}\,)},
\EQN onega2;d \cr}
$$
where $n'$ ($p'$) is the neutral (charged) member of the $N'$ isodoublet. 
In the simple scheme in which we saturate the sum rule with
$\Delta (1232)$ and the $N(1440)$, we find

$$
\ga^2+\ga'^{\;2}=1+{\textstyle{4\over 9}}\ {\cal C}_{\sss\Delta N}^2.
\EQN awsimplesat$$
One can easily find additional relations among these parameters by
taking appropriate matrix elements of \Eq{aw} and 
constructing the Adler-Weisberger sum rules 
for $\pi \Delta$ and $\pi N'$ scattering. However, as we will see next,
it is far more general and practical to work directly with the representations of the
chiral symmetry group.

\vskip0.1in
\noindent {\twelvepoint{\bf 3.\quad The Chiral Representation Theory}}
\vskip0.1in

\vskip0.1in
\noindent {\twelvepoint{\it 3.1\quad Axial-Vector Couplings}}

\noindent Any consequence of the chiral algebra can be obtained from the group
representation theory. We will do so now, as it proves to be a much more
powerful means of extracting consequences of chiral symmetry.
The massless flavors of the underlying QCD Lagrangian transform as $(\oneht,0)$
and $(0,\oneht )$ with respect to $SU(2)_L\times SU(2)_R$.  Baryons made of
light flavors are in general {\it reducible} sums of any number of irreducible
representations, constrained only by isospin.  For $N_c=3$, the allowed
irreducible representations are $(\oneht, 0)$, $(0, \oneht )$, $(\threeht,0)$,
$(0,\threeht )$, $(1,\oneht)$ and $(\oneht ,1)$~$^4$\vfootnote4{\tenpoint These
representations correspond to the dimensionalities, $({\bf 2},{\bf 1})$, 
$({\bf 1},{\bf 2} )$, $({\bf 4},{\bf 1})$, $({\bf 1},{\bf 4} )$, $({\bf 3},{\bf 2})$ and 
$({\bf 2},{\bf 3})$, respectively.}. Since $(\oneht ,1)$ contains isospin $\oneht$ and
$\threeht$, we will differentiate the isospin states by a subscript.  The most
general nucleon and $\Delta$ states with $\lambda=\oneht$ can be written as

$$\EQNalign{& \ket{N,\oneht}= \sum_{\sss l}\,a_{\sss l}\, {\ket{\oneht ,0 }^{\sss (l)}}+
           \sum_{\sss k}\,b_{\sss k}\, {\ket{0, \oneht }^{\sss (k)}}+
           \sum_{\sss m}\,c_{\sss m}\, {\ket{\oneht ,1 }_{\onehtss}^{\sss (m)}}+
           \sum_{\sss n}\,d_{\sss n}\, {\ket{1, \oneht }_{\onehtss}^{\sss (n)}},
\EQN mostgeneralnandd;a \cr
& \ket{\Delta,\oneht}= \sum_{\sss l}\,e_{\sss l}\, {\ket{\threeht ,0 }^{\sss (l)}}+
           \sum_{\sss k}\,f_{\sss k}\, {\ket{0, \threeht }^{\sss (k)}}+
           \sum_{\sss m}\,g_{\sss m}\, {\ket{\oneht ,1 }_{\threehtss}^{\sss (m)}}+
           \sum_{\sss n}\,h_{\sss n}\, {\ket{1, \oneht }_{\threehtss}^{\sss (n)}},
\EQN mostgeneralnandd;b}
$$
where $a_{\sss l}$, ..., $h_{\sss n}$ 
are {\it a priori} unknown mixing parameters.
Parity conservation implies that the $\lambda =-\oneht$ representation is
obtained from \Eq{mostgeneralnandd} by interchanging $SU(2)_L$ and $SU(2)_R$ 
representations; i.e. $\ket{{\cal A},{\cal B}}\rightarrow\ket{{\cal B},{\cal A}}$
where ${\cal A}\in SU(2)_L$ and ${\cal B}\in SU(2)_R$\ref{wein1}. It is
important to realize that
the chiral multiplet structure in the broken chiral symmetry phase has nothing
to do with parity doubling. We define the axial-vector couplings of the $\lambda=\oneht$
nucleon as 

$$\EQNalign{& \bra{N,\oneht}{{\cal Q}^{\sss A}_a}\ket{N,\oneht}={\ga^{({\sss 1/2})}}\,[{T_{a}}]_{\onehtss\onehtss} \ ,
\EQN axialsdef;a \cr
& \bra{N,\oneht}{{\cal Q}^{\sss A}_a}\ket{\Delta,\oneht}=\sqrt{2\over 3}\,{{\cal C}_{\sss\Delta N}^{({\sss 1/2})}}\,[{T_{a}}]_{\onehtss\threehtss}
\EQN axialsdef;b \cr}
$$
where ${{\cal Q}^{\sss A}_a}$ is an $SU(2)_{\sss A}$ generator
and $[{T_{a}}]_{\sss\alpha\beta}$ is the matrix element of the isospin operator
between baryon states of isospin $\alpha$ and $\beta$. Notice that the
axial-vector coupling carries a helicity superscript.

The action of the QCD generators on the states of 
definite chirality is given by

$$\EQNalign{& \;\;\bra{0, \oneht }\,{{\cal Q}^{\sss A}_{a}}\,\ket{0, \oneht }=
-\bra{\oneht ,0 }\,{{\cal Q}^{\sss A}_{a}}\,\ket{\oneht ,0 }=
-[{T_{a}}]_{\onehtss\onehtss} \ ,
\EQN actionrep;a \cr
&{}_{\onehtss}\bra{\oneht ,1}\, {{\cal Q}^{\sss A}_{a}}\,\ket{\oneht ,1}_{\onehtss}=
-{}_{\onehtss}\bra{1,\oneht}\, {{\cal Q}^{\sss A}_{a}}\,\ket{1,\oneht}_{\onehtss}=
-{\textstyle{5\over 3}}[{T_{a}}]_{\onehtss\onehtss} \ ,
\EQN actionrep;b \cr
&{}_{\onehtss}\bra{\oneht ,1}\, {{\cal Q}^{\sss A}_{a}}\,\ket{\oneht ,1}_{\threehtss}=
-{}_{\onehtss}\bra{1,\oneht}\, {{\cal Q}^{\sss A}_{a}}\,\ket{1,\oneht}_{\threehtss}=
2{\sqrt{2\over 3}}[{T_{a}}]_{\onehtss\threehtss}.
\EQN actionrep;c}
$$ 

Using \Eq{mostgeneralnandd} we can now find expressions
for the axial couplings in terms of the coefficients
$a_{\sss l}$, ..., $h_{\sss n}$.
For example, putting \Eq{mostgeneralnandd;a} into \Eq{axialsdef}
and using \Eq{actionrep},
we find for $\lambda=+\oneht$

$${\ga}^{({\sss 1/2})}=
{{\sum_{\sss l}\,|a_{\sss l}|^2-
\sum_{\sss k}\,|b_{\sss k}|^2-
{\textstyle{5\over 3}}(\sum_{\sss m}\,|c_{\sss m}|^2-
\sum_{\sss n}\,|d_{\sss n}|^2)}\over
{\sum_{\sss l}\,|a_{\sss l}|^2+
\sum_{\sss k}\,|b_{\sss k}|^2+
\sum_{\sss m}\,|c_{\sss m}|^2+
\sum_{\sss n}\,|d_{\sss n}|^2}},
\EQN mostgeneralga$$
with the opposite sign holding for $\ga^{({\sss -1/2})}$.
Similar relations can be derived for other axial couplings. We see from this
expression that the baryon axial couplings are completely determined by the
angles which mix the states of definite chirality.

\vskip0.1in
\noindent {\twelvepoint{\it 3.2\quad The Mass-Squared Matrix}}

In order to see the effects of spontaneous symmetry breaking, 
we must consider the baryon mass-squared matrix. 
How does the mass-squared matrix transform with respect to $SU(2)_L\times SU(2)_R$?
In principle the mass-squared matrix, ${\hat M}^2$, can transform as a
singlet plus any non-trivial representation(s) of the chiral group.  Here
we will assume 

$${\hat M}^2={\hat M}_0^2+{\hat M}_{\bar q q}^2,
\EQN masssquared
$$ 
where ${\hat M}_0^2\in (0,0)$ and ${\hat M}_{\bar qq}^2\in(\oneht,\oneht)$. 
In \Ref{alg}, Weinberg showed that the assumption that the forward
$\pi N$ amplitude with $I=2$ in the $t$-channel satisfies a
superconvergence relation is equivalent to the algebraic
statement of \Eq{masssquared} in helicity-conserving frames\ref{wein1}.  
All chiral symmetry breaking in the baryon sector is then contained in 
matrix elements of the form $\bra{0,\oneht}\,{{\hat M}_{\bar q
q}^2}\,\ket{\oneht ,1}$ and 
$\bra{0,\oneht}\,{{\hat M}_{\bar q q}^2}\,\ket{\oneht ,0}$.
The mass of a baryon $B$ is defined as

$$
{M_{\sss B}^2}= \bra{B,\oneht}{{\hat M}^2}\ket{B,\oneht}.
\EQN massdefined
$$
If one assumes that there is no inelastic diffractive
scattering\ref{wein1}, there is an additional superconvergence relation which can be
expressed algebraically as

\offparens
$$\left[\ {\hat M}_0^2\ ,\ {\hat M}_{\bar q q}^2\ \right]\ =\ 0 \ .
\EQN massdiff
$$ 
This rather peculiar commutator constrains the mixing angles in reducible representations, as we will see below.

\vskip0.1in
\noindent {\twelvepoint{\bf 4.\quad Chiral Representations}}
\vskip0.1in

In principle the baryon representations can be
infinite dimensional. If this were the case, it would be unlikely
that chiral symmetry would have any predictive power
for the baryons.
Fortunately, the phenomenological analysis of the Adler-Weisberger relation 
presented in Sect. 2 suggests that the baryon representations 
are small. In this section we will consider the consequences of the simplest 
baryon representations. We will be primarily concerned
with the pion transitions of the ground-state nucleon.
Therefore, it is sufficient to consider the $\lambda = \pm \oneht$
sector, although other helicities can be considered as well.
We will ignore the overall phases of the axial couplings. They can easily
be found using \Eq{axialsdef} and \Eq{actionrep}.

\vskip0.1in
\noindent {\twelvepoint{\it 4.1\quad The Sigma Model}}

\noindent 

The simplest chiral representation places the nucleon helicity states in the
$(\oneht, 0)$ or $(0, \oneht )$ representations.
This corresponds to
$a_l=c_m= d_n=0$ and $b_k =\delta_{k1}$ 
in \Eq{mostgeneralnandd;a}:

$$\ket{N,\oneht}=\ket{0, \oneht }\ \ , \ \ \ket{N,-\oneht}=\ket{\oneht ,0}\ \ .
$$
{}From \Eq{mostgeneralga} we find immediately that $\ga =1$. 
This prediction follows exclusively from chiral symmetry
and the assumed representation. This result is familiar from
the linear sigma model where there is one nucleon field,
$N=N_L+N_R$, with $N_L\in(\oneht, 0)$ and
$N_R\in(0,\oneht)$, and one finds $\ga =1$.
Here we see that helicity and chirality are identified in the collinear frame.
\vskip0.1in
\noindent {\twelvepoint{\it 4.2\quad The Generalized Sigma Model}}

\noindent 
A generalized sigma model contains nucleons and other $I={1\over 2}$ baryons
which are sums of 
any number of $(\oneht, 0)$ and $(0, \oneht )$ representations; that is,
models where $c_m= d_n=0$ in \Eq{mostgeneralnandd;a}.
These models are unrealistic as $\ga$ cannot exceed unity. 
This is easily seen from \Eq{mostgeneralga}:

$$|\ga|=\Big|
{{\sum_{\sss l}\,|a_{\sss l}|^2-
\sum_{\sss k}\,|b_{\sss k}|^2}\over
{\sum_{\sss l}\,|a_{\sss l}|^2+
\sum_{\sss k}\,|b_{\sss k}|^2}}\Big|\leq 1 \ .
\EQN mostgeneralgawods$$
Hence, it is not possible to reach the physical value of 
$\ga$ unless the nucleon couples to at least one state which 
transforms as $(1,\oneht)$ or $(\oneht ,1)$.
Put another way, unless the nucleon couples to $\Delta$, or some other
$I=\threeht$ state, $\ga$ is constrained by the bound in \Eq{mostgeneralgawods}.

\vskip0.1in
\noindent {\twelvepoint{\it 4.3\quad The Non-Relativistic Quark Model}}

\noindent 
The simplest realistic representation places $N$ and $\Delta$
in an irreducible $(\oneht ,1 )$ representation:

$$\EQNalign{
& \ket{N,{\oneht}}= \ket{\oneht ,1 }_{\onehtss},
\EQN nandd;a \cr
& \ket{\Delta,\oneht}= \ket{\oneht ,1 }_{\threehtss}.
\EQN nandd;b}
$$
That is, $a_l=b_k = d_n= e_l=f_k = h_n=0$ and $c_m= g_m=\delta_{m1}$ 
in \Eq{mostgeneralnandd;a}. This representation is interesting
because it gives results equivalent to the spin-flavor $SU(4)$ predictions of the NCQM\ref{strong}\ref{beane2}.
For the couplings and masses we find

$$
\ga = \frac{5}{3}\ \ ,\ \ {\cal C}_{\sss\Delta N}= 2\ \ ,\ \ M^2_N=M^2_\Delta \ \ .
\EQN nanddcoupl
$$
This representation is unrealistic since the $\Delta$ and the nucleon are degenerate.
In extensions of the NCQM one effectively perturbs around this basis in order
to split the nucleon and the $\Delta$ and quench the axial couplings.
\vskip0.1in
\noindent {\twelvepoint{\it 4.4\quad A Minimal Realistic Model}}

\noindent As we have seen above, a more interesting and realistic scenario is one in which $N$,
$\Delta$ and $N'$ saturate the Adler-Weisberger sum
rule. One can easily show that the unique solution in which these three states 
communicate by pion emission and absorption is a reducible sum 
$(0, \oneht )\oplus(\oneht ,1)$\ref{Fred}\ref{wein1}. For $\lambda =\oneht$ 
we can write this representation in terms of a single mixing angle
$\theta$ as

$$\EQNalign{& \ket{N,\oneht}=\sin\theta\,\ket{0, \oneht }+\cos\theta\,\ket{\oneht ,1}_{\onehtss},
\EQN mainrep;a \cr
& \ket{N',\oneht}=-\cos\theta\,\ket{0, \oneht }+\sin\theta\,\ket{\oneht ,1}_{\onehtss},
\EQN mainrep;b \cr
& \ket{\Delta,\oneht}=\ket{\oneht ,1}_{\threehtss}.
\EQN mainrep;c}
$$

{}From \Eq{mainrep} and \Eq{actionrep} we immediately find 

$$\EQNalign{
& \ga =1+\textstyle{2\over 3}\cos^2\theta \ ,
\EQN param;a \cr
& \ga' =\textstyle{2\over 3}\sin\theta\cos\theta \ ,
\EQN param;b \cr
& \ga'' =1+\textstyle{2\over 3}\sin^2\theta \ ,
\EQN param;c \cr
&{\cal C}_{\sss\Delta N}=2\cos\theta \ ,
\EQN param;d \cr
&{\cal C}_{\sss\Delta N'}=2\sin\theta \ ,
\EQN param;e \cr}
$$
which clearly is consistent with \Eq{awsimplesat}, the Adler-Weisberger sum rule for $\pi
N$ scattering. It is straightforward to show that there is a single relation involving
the three masses and the mixing angle, $\theta$:

$$
\cos^2\theta\, {M_{\sss N}^2}+\sin^2\theta\, {M_{\sss N'}^2}
={M_{\sss\Delta}^2} \ .
\EQN massrelation
$$
The constraint of no inelastic diffraction from \Eq{massdiff} implies
maximal mixing, $\theta= 45^o$, which results in $\ga=1.33$ and $M_{N'}=1467~{\rm MeV}$.
Other predictions are shown in Table 2. This value of the nucleon axial
coupling is consistent with its chiral limit value, as determined by the
process $\pi N\rightarrow \pi\pi N$~\ref{Fettes}.
We have also given predictions that
follow from fitting the mixing angle to $\ga=1.26$. The error bars on the
experimental values correspond
to the PDG ranges for the decay widths and branching fractions and therefore are seriously underestimated, particularly for
the Roper axial couplings. Notice that this chiral representation accounts for
the Roper mass while quenching the nucleon axial couplings from the NCQM values toward
the experimental values. We find this to be compelling evidence that this
chiral representation is perturbatively close to nature. Since the $N$-$N'$
mass splitting is of order the kaon mass, the chiral (continuum) corrections to
this chiral multiplet can be computed using chiral perturbation theory\ref{beanesav}.

\table{gl}
\tenpoint
\caption{\twelvepoint Comparison of chiral predictions with experiment. 
In the second column (TH1) we fit $\theta =51^o$ from $\ga=1.26$, the physical value of the axial coupling.
In the third column (TH2) we assume maximal mixing $\theta =45^o$, consistent
with the constraint of no inelastic diffraction.}
\doublespaced
\ruledtable
{}    | ${\rm TH1}$ | ${\rm TH2}$  | ${\rm EXP}$ | ${\rm PROCESS}$  \cr
$M_{\sss N'}$   | $1386$ | $1467$ |  $1440\pm30$ |  $N\pi\rightarrow N\pi ,\ldots$ \cr  
$\ga$   | $1.26$ (input) | $1.33$  | $1.26$ |  $N\rightarrow N\pi$ \cr  
$\ga'$   | $0.33$ | $0.33$  | $ 0.71 \pm 0.20$ |  $N'\rightarrow N\pi$ \cr  
$\ga''$   | $1.41$ | $1.33$  | $--$ |  $N'\rightarrow N'\pi$ \cr  
${\cal C}_{\sss\Delta N}$  | $1.25$ | $1.41$  | $1.51 \pm 0.10$ |  $\Delta\rightarrow N\pi$ \cr  
${\cal C}_{\sss\Delta N'}$  | $1.56$ | $1.41$  | $1.38 \pm 0.50$ |  $N'\rightarrow \Delta\pi$ 
\endruledtable
\endtable

\vskip0.1in
\noindent {\twelvepoint{\it 4.5\quad A Second Reducible Model}}

\noindent A second interesting scenario is one in which $N$,
$\Delta$, $N'$ and $\Delta'$ ($\Delta (1600)$) saturate the Adler-Weisberger sum
rule. Here there are several ways in which these states can be embedded
in the chiral algebra. Here we choose a reducible sum of $(1, \oneht )$ and $(\oneht ,1)$. 
For $\lambda =\oneht$ we can write this representation in terms of two mixing angles,
$\phi$ and $\delta$ as

$$\EQNalign{& \ket{N,\oneht}=\sin\phi\,\ket{\oneht ,1}_{\onehtss}'+\cos\phi\,\ket{1,\oneht}_{\onehtss},
\EQN mainrep3;a \cr
& \ket{N',\oneht}=-\cos\phi\,\ket{\oneht ,1}_{\onehtss}'+\sin\phi\,\ket{1,\oneht}_{\onehtss},
\EQN mainrep3;b \cr
& \ket{\Delta,\oneht}=\sin\delta\,\ket{\oneht ,1}_{\threehtss}'+\cos\delta\,\ket{1,\oneht}_{\threehtss},
\EQN mainrep3;c \cr
& \ket{\Delta',\oneht}=-\cos\delta\,\ket{\oneht ,1}_{\threehtss}'+\sin\delta\,\ket{1,\oneht}_{\threehtss}.
\EQN mainrep3;d}
$$
{}From \Eq{mainrep} and \Eq{actionrep} we immediately find 

$$\EQNalign{
& \ga =\textstyle{5\over 3}\cos 2\phi ,
\EQN param3;a \cr
& \ga' =\textstyle{5\over 3}\sin 2\phi ,
\EQN param3;b \cr
& \ga'' =\textstyle{5\over 3}\cos 2\phi ,
\EQN param3;c \cr
&{\cal C}_{\sss\Delta N}={\cal C}_{\sss\Delta'N'}=2\cos (\phi +\delta ),
\EQN param3;d \cr
&{\cal C}_{\sss\Delta N'}={\cal C}_{\sss\Delta'N}=2\sin (\phi +\delta ).
\EQN param3;e \cr}
$$
It is straightforward to show that there is a single relation that is
independent of the mixing angles, and one relation involving
the mixing angles:

$$\EQNalign{
{M_{\sss N}^2}+{M_{\sss N'}^2}&={M_{\sss\Delta}^2} +{M_{\sss\Delta'}^2},
\EQN massrelations3;a \cr
({M_{\sss\Delta'}^2}-{M_{\sss\Delta}^2})\cos{2\delta}&= ({M_{\sss
N'}^2}-{M_{\sss N}^2})\cos{2\phi} .
\EQN massrelations3;b \cr}
$$
With $\ga$ as input, \Eq{param3} determines $\phi =21^o$.
Using the masses of $N$, $\Delta$ and $\Delta'$ as input in
\Eq{massrelations3;a} gives $M_{N'}=1790~{\rm MeV}$. We then find
that \Eq{massrelations3;b} has no real solution for $\delta$. We therefore
find that this reducible representation is not consistent with the assumed
particle content. We will return to this point below.

\vskip0.1in
\noindent {\twelvepoint{\bf 5.\quad Discussion}}
\vskip0.1in

\vskip0.1in
\noindent {\twelvepoint{\it 5.1\quad The Quark Model Interpretation}}

\noindent The spin-flavor structure of the baryon multiplets seems to provide a
powerful explanation of why the Adler-Weisberger sum rule is almost completely
saturated by the $\Delta$, with smaller contributions from higher states. 
In the NCQM the nucleon and the $\Delta$ resonance fill out the
completely symmetric ${\bf 20}$-dimensional representation of spin-flavor
$SU(4)$, which we have seen is equivalent to the 
$(\oneht ,1 )$ representation of $SU(2)_L\times SU(2)_R$\ref{strong}\ref{beane2}.
In the NCQM the proton and $\Delta^+$ wavefunctions can be written as

$$\EQNalign{
&\ket{\;p\; ;\;{\bf 20}}={1\over\sqrt{6}}  (2\ket{u\uparrow u\uparrow d\downarrow}
-\ket{u\uparrow u\downarrow d\uparrow} - \ket{u\downarrow u\uparrow d\uparrow} )\; +\ldots
\EQN protwf56;a \cr
&\ket{\;\Delta^+\; ;\;{\bf 20}}={1\over\sqrt{3}}  (\ket{u\uparrow u\uparrow d\downarrow}
+\ket{u\uparrow u\downarrow d\uparrow} + \ket{u\downarrow u\uparrow d\uparrow} )\; +\ldots
\EQN protwf56;b \cr}
$$
where $\ldots$ signifies cyclic permutations which are irrelevant for our purpose.
The action of the axial-vector operator, $q^\dagger \sigma^3\tau^3 q$, on
$u\uparrow$ and $d\downarrow$ is $+1$ and on
$u\downarrow$ and $d\uparrow$ is $-1$. One then trivially finds $\ga =5/3$ and
${\cal C}_{\sss\Delta N}=2$. Similarly, placing the proton in the completely antisymmetric
${\bf 4}$ representation gives rise to the wavefunction

$$
\ket{\;p\; ;\;{\bf 4}}={1\over\sqrt{2}} (\ket{u\uparrow u\downarrow d\uparrow}
-\ket{u\downarrow u\uparrow d\uparrow})\; +\ldots
\EQN protwf20
$$
from which one easily finds $\ga =1$. We can then easily recover the
axial-coupling predictions from our minimal realistic model by placing $N$, $N'$
and $\Delta$ is a reducible ${\bf 4}\oplus{\bf 20}$ representation of
$SU(4)$:

$$\EQNalign{& \ket{N}=\sin\theta\,\ket{\;{\bf 4}\; ;\;
1^+}+\cos\theta\,\ket{\;{\bf 20}\; ;0^+}_{4},
\EQN qmmainrep;a \cr
& \ket{N'}=-\cos\theta\,\ket{\;{\bf 4}\; ;\; 1^+}+\sin\theta\,\ket{\;{\bf 20}\;
;0^+}_{4},
\EQN qmmainrep;b \cr
& \ket{\Delta}=\ket{\;{\bf 20}\; ;0^+}_{16} \ ,
\EQN qmmainrep;c}
$$
where the subscripts indicate the spin-flavor content.
Here we have included the spatial quantum numbers that one naively expects. Since the
${\bf 4}$ of spin-flavor $SU(4)$ is completely antisymmetric, it must carry at least one
unit of orbital angular momentum. In the NCQM the $\ket{\;{\bf 4}\; ;\; 1^+}$
state is thought to be irrelevant as it requires two quarks in a baryon
to be in an excited state. The presence of orbital angular momentum is quite
strange as a nonvanishing nucleon-$\Delta$ mass splitting requires that
${\hat M}_{\bar q q}^2$, which acts like an order parameter, carry
orbital angular momentum. The peculiar NCQM interpretation of the chiral
symmetry representations in the collinear frame was noticed long ago by
Casher and Susskind\ref{Casher}.

In the NCQM one usually assigns $N'$ and $\Delta'$ to a radially-excited
${\bf 20}$-dimensional representation of $SU(4)$. These states 
then mix with the ``ground state'' ${\bf 20}$-dimensional representation containing
$N$ and $\Delta$. But this is precisely the second reducible model that
we have analyzed above which overpredicts the Roper mass and has no solution
for the axial couplings.

\vskip0.1in
\noindent {\twelvepoint{\it 5.2\quad The Large-$\nc$ Limit}}

Given the peculiar interpretation of the minimal realistic model in
the NCQM it is of interest to consider the large-$\nc$ limit. For a baryon made of
$\nc$ (odd) quarks, there are $(\nc +1)(\nc +3)/4$ possible irreducible
representations.  It can be shown that, for each helicity $\lambda$, the ground
state baryons fall into an $(({N_c}- 2\lambda )/4,({N_c}+ 2\lambda)/4)$
irreducible representation of $SU(2)_L\times SU(2)_R$\ref{beane2}. It is then
straightforward to find

$$\EQNalign{
&\ga={{{(\nc +2)}/3}},
\EQN onega;a \cr
&{\cal C}_{\sss\Delta N}=\sqrt{{(\nc +5)(\nc -1)}}/2,
\EQN onega;b \cr}
$$
which is consistent with the contracted spin-flavor $SU(4)$ symmetry of large-$\nc$
QCD\ref{Manohar:1998xv} and for $\nc =3$ is consistent with the $SU(4)$ results.

We may now ask whether our minimal realistic model has a sensible
large-$\nc$ limit. If we assume that a baryon is made of $\nc$ quarks then our
model generalizes to $(({N_c}- 1)/4,({N_c}+ 1)/4)\oplus (0,{\oneht })$, again
with a single mixing angle, $\theta$.  Since $M_{\sss\Delta}-M_{\sss
N}=O(1/\nc)$, it follows from \Eq{massrelation} that
$\theta=O(1/\nc)$. Therefore in the large-$\nc$ limit $N'$ decouples from the
reducible chiral representation and transforms irreducibly in the 
$(0,{\oneht })$ representation (the ${\bf 4}$ of $SU(4)$), and $N$ and $\Delta$ transform
irreducibly in the $(({N_c}- 1)/4,({N_c}+ 1)/4)$ representation as
required. Hence the minimal realistic representation is consistent with
large-$\nc$ expectations for the ground state tower.  What is surprising is
that data seem to prefer maximal mixing, $\sin\theta\sim\cos\theta$, whereas in
the large-$\nc$ limit, $\cos\theta\sim 1$ and $\sin\theta\sim{1/\nc}$. We do
not yet understand what this portends for the ${1/\nc}$ corrections of the
nucleon and $\Delta$ axial couplings.

\vskip0.1in
\noindent {\twelvepoint{\bf 6.\quad Conclusion}}
\vskip0.1in

\noindent The main point to take from this paper is that even at low energies
where chiral symmetry is spontaneously broken, there is a sense in which the
baryon spectrum fall into {\it reducible} representations of the chiral
algebra. This has nothing to do with parity doubling near a
chiral symmetry restoring phase transition. We have found that existing data
suggest that the nucleon and the $\Delta$ and Roper resonances form a reducible
sum of $(0, \oneht )$ and $(\oneht ,1)$ representations of the chiral group,
with maximal mixing. From the perspective of the naive constituent quark model
this is equivalent to placing these states in a reducible 
${\bf 4}\oplus{\bf 20}$ representation of
spin-flavor $SU(4)$. Our results suggest that other baryons also fall into
finite-dimensional chiral representations that in principle can be mapped out
at JLab and other experimental facilities.
We stress that it is somewhat peculiar that the chiral multiplet involving the
nucleon involves only a few states and that the representations enter with
approximately equal weight\ref{beanesav}. We find no QCD-based argument which would explain
this simple multiplet structure. This is a worthy puzzle whose resolution --we
believe-- will lead to deep insight into the manner in which the hadron
spectrum arises from QCD.

\vskip0.1in
\noindent {\twelvepoint{\bf Acknowledgments}}
\vskip0.1in

\noindent We thank Martin Savage and Ishmail Zahed for valuable conversations. 
This research was supported in part by the 
DOE grant DE-FG03-00-ER-41132 (SRB), 
and by RIKEN, Brookhaven National Laboratory,
the DOE grant DE-AC02-98CH10886, OJI award DE-FG03-01ER41196 
and by an Alfred P.~Sloan Fellowship (UvK).


\nosechead{References}
\ListReferences \vfill\supereject \end